# Superlattices of $Bi_2Se_3$/$In_2Se_3$: Growth Characteristics and Structural Properties

Z. Y. Wang[1], X. Guo[1], H. D. Li[1,2], T. L. Wong[3], N. Wang[3], M. H. Xie[1,a)]

[1]*Physics Department, The University of Hong Kong, Pokfulam Road, Hong Kong, China*
[2]*Department of Physics, Beijing Jiaotong University, Beijing 100044, China*
[3]*Physics Department, Hong Kong University of Science and Technology, Clear Water Bay, Kowloon, Hong Kong, China*

***Abstract:***

Superlattices (SLs) consisted of alternating $Bi_2Se_3$ and $In_2Se_3$ layers are grown on Si(111) by molecular-beam epitaxy. $Bi_2Se_3$, a three-dimensional topological insulator (TI), showed good chemical and structural compatibility with $In_2Se_3$, a normal band insulator with large energy bandgap. The individual layers in the SLs are very uniform and the hetero-interfaces are sharp. Therefore, such SL structures are potential candidates for explorations of the quantum size effects of TIs.

[a)]Author to whom correspondence should be addressed. Email: mhxie@hku.hk

Along with the extensive researches of materials and properties of three- dimensional (3D) topological insulators (TIs),[1] attention has increasingly been paid on ultrathin films and nanostructures of such materials for enhanced effects and properties associated with the topological states of electrons.[2] In the same line of thoughts, multi-layered structures constituted of TIs and normal band insulators, such as superlattices (SLs) or multiple quantum well (MQWs) of …Bi2Se3/ZnSe…, have been attempted by the technique of molecular-beam epitaxy (MBE).[3] Since SLs and MQWs can be used for future device applications, successful growth of high quality SL or MQW samples is of great fundamental and practical importance. Unfortunately, the combination of topological insulator $Bi_2Se_3$ and normal band insulator ZnSe does not lead to desired heterostructures with good structural quality.[3] Two complementary interfaces, i.e., $Bi_2Se_3$-on-ZnSe and ZnSe-on-$Bi_2Se_3$, are asymmetric in morphology and strain state. The structures are thus not very uniform and do not meet the requirements for investigations of the quantum size effect and for device applications.

In this Letter, we report successful growth of high quality …$Bi_2Se_3$/$In_2Se_3$… SL structures with great structural quality. The interfaces of $Bi_2Se_3$/$In_2Se_3$ and $In_2Se_3$/$Bi_2Se_3$ are symmetrical and sharp. Transmission electron microscopy (TEM) examinations of the samples reveal few defects in samples, indicating the strain relaxation is at the van der Waals gaps of the hetero-interfaces. X-ray diffraction (XRD) measurements show satellite peaks characteristic of the SL structure of the samples, suggesting good uniformity of the individual layers.

The compound of $In_2Se_3$ is a well known semiconductor (insulator) that has an energy bandgap of 1.2-1.3 eV.[4,5] It is conveniently grown in MBE reactors designated for $Bi_2Se_3$ growth, as both are selenide compounds, showing similar growth conditions. $In_2Se_3$ is also a layered material as of $Bi_2Se_3$, yet it has a relatively small lattice mismatch with $Bi_2Se_3$ (~ 3.4%). Therefore, the two are chemically and structurally compatible, suitable for growth of the SL or MQW structures.

The growth experiments and surface characterizations of the heterostructures are made in a customized MBE system, where indium (In), bismuth (Bi) and selenium (Se) fluxes are generated from Knudsen cells.[6] In situ reflection high energy electron diffraction (RHEED) is employed to monitor the growing surfaces in real-time to assess the film growth mode, lattice misfit strain and surface reconstructions. The RHEED specular-beam intensity is recorded, and from its oscillations we obtain the deposition rates of the films.[6] Room temperature (RT) scanning tunneling microscopy (STM) is used to examine the morphologies of the grown samples, where the tunneling current is 0.2 nA and the sample bias is -0.45 V. The substrates are nominally flat Si(111), which are deoxidized at ≥1000 ºC in vacuum for clear (7 × 7) surfaces as evidenced by the RHEED. Afterwards, we prepare a thin InSe buffer on Si(111)- (7 × 7) by firstly depositing about 3 monolayers In at ~ 220 K, followed by annealing in a flux of Se at 490K until a set of (√3 × √3)*R*30° pattern appears in the RHEED. The distance between the main diffraction streaks is found to change by about 5% from that of Si, suggesting a lattice constant of ~ 4 Å. This is consistent with the lattices of the layered α-phase $In_2Se_3$. On the other hand, some previous studies showed the (√3 × √3)*R*30° surfaces only for γ-phase $In_2Se_3$.[5,7] However, γ-phase $In_2Se_3$ has a lattice parameter (~ 4.26Å) that is far larger than that measured by the RHEED. Therefore, we tend to believe the buffer is of α-phase $In_2Se_3$. As for the origin of the (√3 × √3)*R*30° superstructure on such a surface, it remains unclear.

$Bi_2Se_3$ deposition on $In_2Se_3$ buffer is conducted at 180 ºC using a Bi to Se flux ratio of 1:10. From the persistence of the streaky RHEED patterns, we infer that two-dimensional layer-by-layer growth of $Bi_2Se_3$ is achieved. Indeed, the RHEED intensity starts to oscillate upon the initiation of $Bi_2Se_3$ deposition [Fig. 1 inset (i)]. A sample containing a thin layer of $Bi_2Se_3$ (9nm) deposited on such $In_2Se_3$ buffer is characterized by XRD, and the result of the reflective $\theta$ - $2\theta$ scan is shown in Fig. 1. It reveals not only the diffraction peaks of the substrate (Si) and $Bi_2Se_3$ epifilm, but also Kiessig fringes with the period corresponding well with film thickness, implying good film uniformity and sharp interfaces. No $In_2Se_3$-related peak is seen in the XRD data, however, probably because the buffer is too thin, or they overlap those of $Bi_2Se_3$ film. The inset (ii) in Fig. 1 presents a STM image of the surface after about 1.5 quintuple layers (QLs) $Bi_2Se_3$ deposition on $In_2Se_3$ buffer, revealing an atomically flat surface with 1 QL high nucleation islands. Here, we would like to further mention that such $Bi_2Se_3$ films grown on the $In_2Se_3$ buffers seem to be superior to those grown on the amorphous buffers reported earlier.[3,6] They show similarly good surfaces and electrical properties, if not better, but also improved film adherence to the substrate for easier device processing. A comparison of $Bi_2Se_3$ epifilms grown on different buffers and substrates is summarized in a different publication,[8] and in the following, we focus on the heteroepitaxy of $In_2Se_3$ on $Bi_2Se_3$ and of $Bi_2Se_3$ on $In_2Se_3$ for growth of superlattice structures.

Depositions of $In_2Se_3$ on $Bi_2Se_3$ and of $Bi_2Se_3$ on $In_2Se_3$ are readily achieved by switching on and off the In and Bi fluxes simultaneously while the flux of Se remains unchanged. For the purpose of $In_2Se_3$ growth, the In source has been set at a similar flux to Bi, i.e., In:Se ~ Bi:Se ~ 1:10. Fig. 2a depicts the RHEED specular-beam intensity variation during deposition of the SL structure of alternating $Bi_2Se_3$ and $In_2Se_3$ while typical RHEED patterns of the respective surfaces are shown in the insets. Note that during the first $Bi_2Se_3$ layer deposition on $In_2Se_3$ buffer, the RHEED intensity oscillates as mentioned earlier, but such oscillations disappear later due to a change of the growth mode from island nucleation to step-flow. It is also seen that at the start of $In_2Se_3$ deposition on $Bi_2Se_3$, the RHEED intensity drops suddenly, which may reflect a roughening of the surface. Indeed, a slightly rougher surface of $In_2Se_3$ than that of $Bi_2Se_3$ may be inferred from a thickening and elongation of the diffraction streaks in inset (ii) than (i) of Fig. 2, which are RHEED patterns taken from $In_2Se_3$ and $Bi_2Se_3$ layers, respectively, as indicated. However, we also wish to point out that the specular spot of the RHEED during measurement is in the vicinity of the (009) Bragg spot of bulk $Bi_2Se_3$, therefore a lattice parameter change upon $In_2Se_3$ deposition has made the specular-beam off the Bragg spot, contributing further to the intensity drop. Similarly, upon the commencement of $Bi_2Se_3$ growth on $In_2Se_3$, there is a sharp intensity rise, part of which is due to the smoothening of the surface and another part is due to a shift of the Bragg spot relative to the specular beam. Regardless of the intensity variations, the RHEED pattern remains streaky showing an unstructured $(1 \times 1)$ pattern throughout the deposition process. Therefore, 2D layer-by-layer growth is maintained. This contrasts to the case of $Bi_2Se_3$/ZnSe growth, where there is a tendency of 3D islanding of ZnSe surface when deposited on $Bi_2Se_3$.[3] This fact reflects the very nature of the van der Waals bonding at the $Bi_2Se_3/In_2Se_3$ hetero-interfaces. They are of low surface energy and the lattice misfit strains are easily relieved without breaking chemical bonds.

Strain states of the heterostructures during the SL sample growth are monitored by the RHEED as well. The evolution of the measured reciprocal lattice parameter $D$ (defined in the inset (i) of Fig. 2(a) is shown in Fig. 2(b). Firstly, one observes that the lattices of heteroepitaxial $Bi_2Se_3$ and $In_2Se_3$ films approach their strain-free parameters with increasing deposition thicknesses. A closer look at the strain relaxation processes [Fig. 2(c)] reveals that the residual strain $\varepsilon = (a-a_0)/a_0 = (D_0-D)/D$ follows an exponential relation with time or layer thickness $h$, i.e., $\varepsilon = \varepsilon_0 \exp(-h/\lambda)$,[9,10] where $a$ and $D$ are real- and reciprocal-space lattice parameters, respectively, of the epifilm, while $a_0$ and $D_0$ are the corresponding parameters for a strain-free layer. $\varepsilon_0$ is the lattice misfit between the epifilm and the substrate, and for $Bi_2Se_3/In_2Se_3$, it is ~3.4% as mentioned earlier. The constant $\lambda$ characterizes the rate at which strain relaxes. Least-square fittings of the data in Fig. 2(c) result in $\lambda$ ~ 16 Å and ~ 4 Å, respectively, for the processes of $Bi_2Se_3$ deposition on $In_2Se_3$ and $In_2Se_3$ growth on $Bi_2Se_3$. These are relatively small values (i.e., ≤ 1 ~ 2 QLs), characterizing a fast rate of strain relaxation process. Further, according to elasticity theory, there should exist a critical film thickness $h_c$, below which a coherent film can be grown without strain-relaxation.[11] If so, for a lattice misfit of ~3.4% as in $Bi_2Se_3/In_2Se_3$, $h_c$ ~ 20 Å, assuming the strain-relieving dislocations to have the Burgers vector of ~ 0.4nm in magnitude. This is not what is observed by experiments. Rather, one finds the strain to relax at the very start of the heteroepitaxy. Together with the small $\lambda$ derived above, it asserts a strain relaxation process occurring at the van der Waals interfaces, where $Bi_2Se_3$ and $In_2Se_3$ compounds couple very weakly. Strain relaxation invokes no chemical bond breaking. On the other hand, the fact that strain does not fully relax upon the commencement of hetero-growth indicates some extents of constraints of the lattice by that of the substrate at the van der Waals interface. This conforms to the very nature of van der Waals epitaxy.[6,12]

Lastly in Fig. 2, we observe the surfaces and strain states to evolve highly repeatedly during different periods of …$In_2Se_3/Bi_2Se_3$… SL growth. It implies high uniformity of the structures. To show this, we present, in Fig. 3(a), a XRD $\theta$-$2\theta$ scan of a 20-period …5nm-$In_2Se_3$/10nm-$Bi_2Se_3$… SL sample. For comparison, result from a single-layered heterstructure of $In_2Se_3/Bi_2Se_3$ is also shown. As is seen, up to five satellite peaks are detected from the SL sample, confirming the integrity of the structure. From the period of the satellite peaks as well as from its FFT result [Fig. 3(b)], we derive the period of the SL structure to be of 15 nm, matching exactly with the designed thickness period of the SL structure. Fig. 4 shows a high-resolution TEM micrograph of a similar sample, from which, alternating $In_2Se_3/Bi_2Se_3$ layers are clearly resolvable. The hetero-interfaces are sharp. No extended defect is seen. However, there appear some slight contrast variations, particularly in the regions of $In_2Se_3$ layers. This may be caused by a residual strain field in the film, as deducible from Fig. 2(b) (i.e., the strain is not fully relived at the end of the $In_2Se_3$ layer growth). On the other hand, from the electron diffraction pattern shown in the inset of Fig. 4, we observe diffraction spots from $Bi_2Se_3$ and $In_2Se_3$ crystals (overlapped to each other), suggesting high crystallinity of the phases in the materials.

To conclude, $Bi_2Se_3$ and $In_2Se_3$ form a superior combination for heteroepitaxy of superlattice and/or MQW structures, where topological insulator is embedded in a normal band insulator for exploration of quantum size and surface-coupling effects. The Van der Waals hetero-interface ensures the ideal 2D growth mode of the structures with sharp interfaces. Strain relaxation is realized at the van der Waals interfaces without invoking dislocations. The SL samples of …$In_2Se_3/Bi_2Se_3$… fabricated by MBE show high structural quality, which are potential candidates for future transport and optical studies.

**Acknowledgement**

We wish to thank W.K. Ho and S. Y. Chui for their help in the growth and XRD experiments, respectively. The project is financially supported by a General Research Fund (No. HKU 7061/10P) and a Collaborative Research Fund (HKU 10/CRF/08) from the Research Grant Council of Hong Kong Special Administrative Region.

**Figure Captions:**

Fig. 1 XRD $\theta$-2$\theta$ scan of a 9nm $Bi_2Se_3$ film on $In_2Se_3$ buffer grown on Si(111). Inset (i): RHEED intensity oscillation during the initial stage $Bi_2Se_3$ deposition. Inset (ii) STM image of the surface of a thin (~1.5 QLs) $Bi_2Se_3$ film deposited on the $In_2Se_3$ buffer.

Fig. 2 Evolution of (a) the RHEED specular-beam intensity and (b) the inter-streak spacing ($D$) during deposition of the …$Bi_2Se_3$/$In_2Se_3$… superlattice structure. The inset (i) and (ii) in (a) show the RHEED patterns taken from the growing $Bi_2Se_3$ and $In_2Se_3$ surfaces, respectively. The two dash-dotted lines in figure (b) represent the $D$ values of strain-free $Bi_2Se_3$ and $In_2Se_3$ films obtained from very thick (>100 nm) layers. (c) Derived residual strain $\varepsilon$ from the experimental $D$ in (b), showing the exponential strain-relaxation processes. The thin lines represent results of the least-square fitting of the data.

Fig. 3 (a) XRD $\theta$-2$\theta$ scans of a SL sample (top) and a single-layered heterostructure of $Bi_2Se_3$/$In_2Se_3$ (bottom). (b) FFT of the XRD data of the SL sample, where the downward arrows point to peaks corresponding to the periodicity of the satellite peaks seen in the $\theta$-2$\theta$ scan. Such peaks or the periodicity of the satellites translate into a SL structure with 15nm thickness period.

Fig. 4 TEM micrograph showing the alternating $Bi_2Se_3$/$In_2Se_3$ layers in the SL sample. Inset: Transmission electron diffraction data of the sample.

# Fig.1

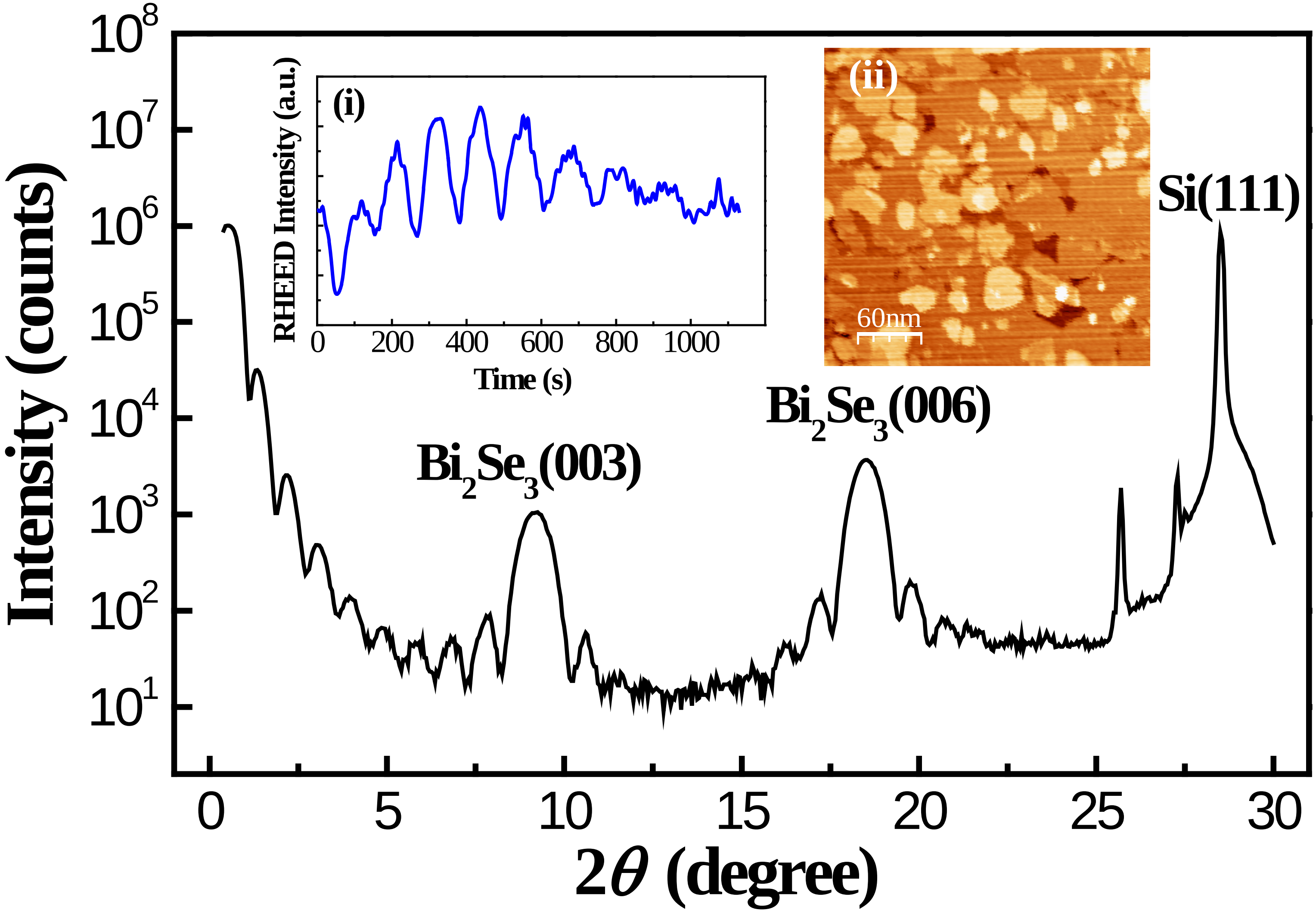

# Fig.2

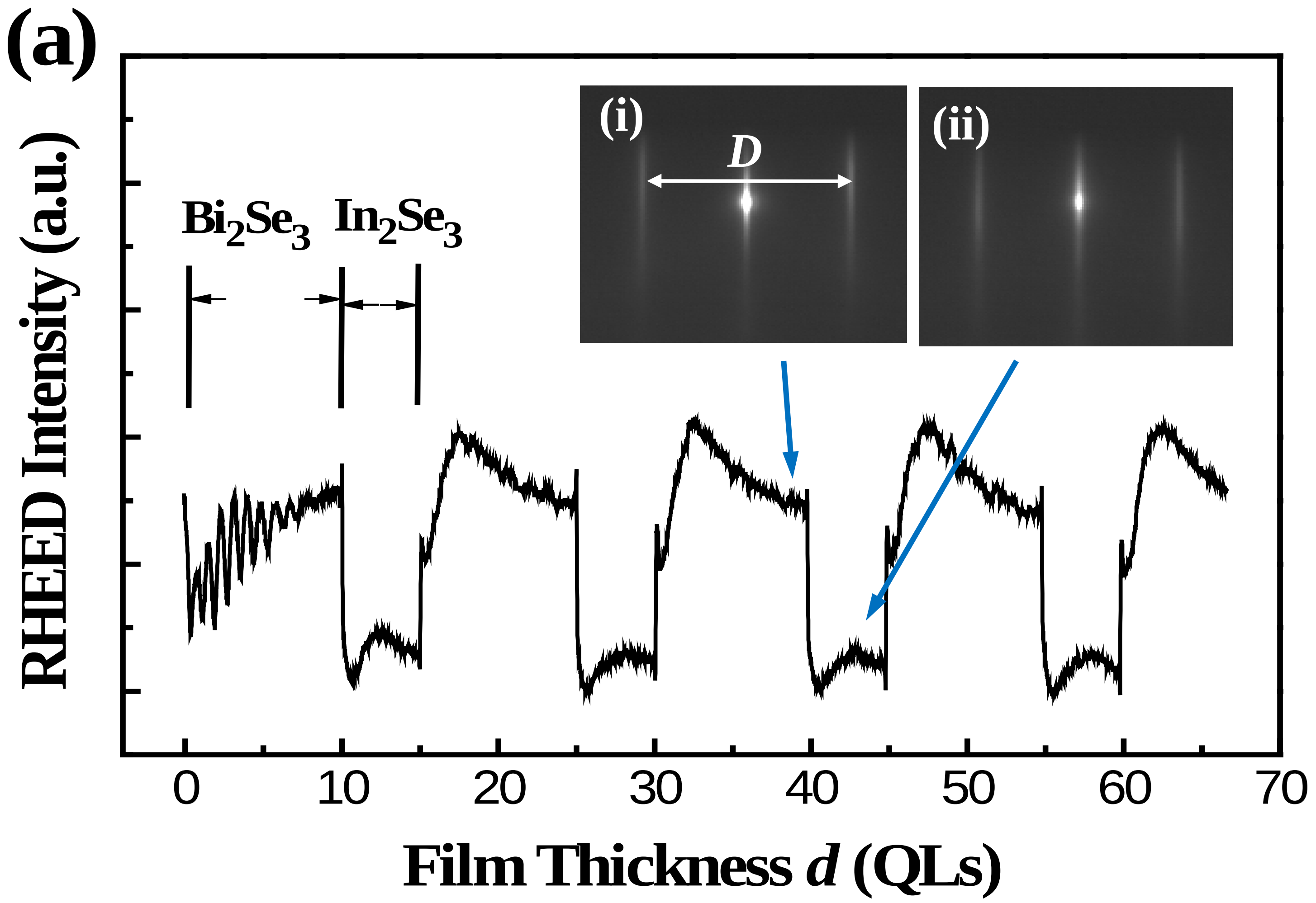

# Fig 2

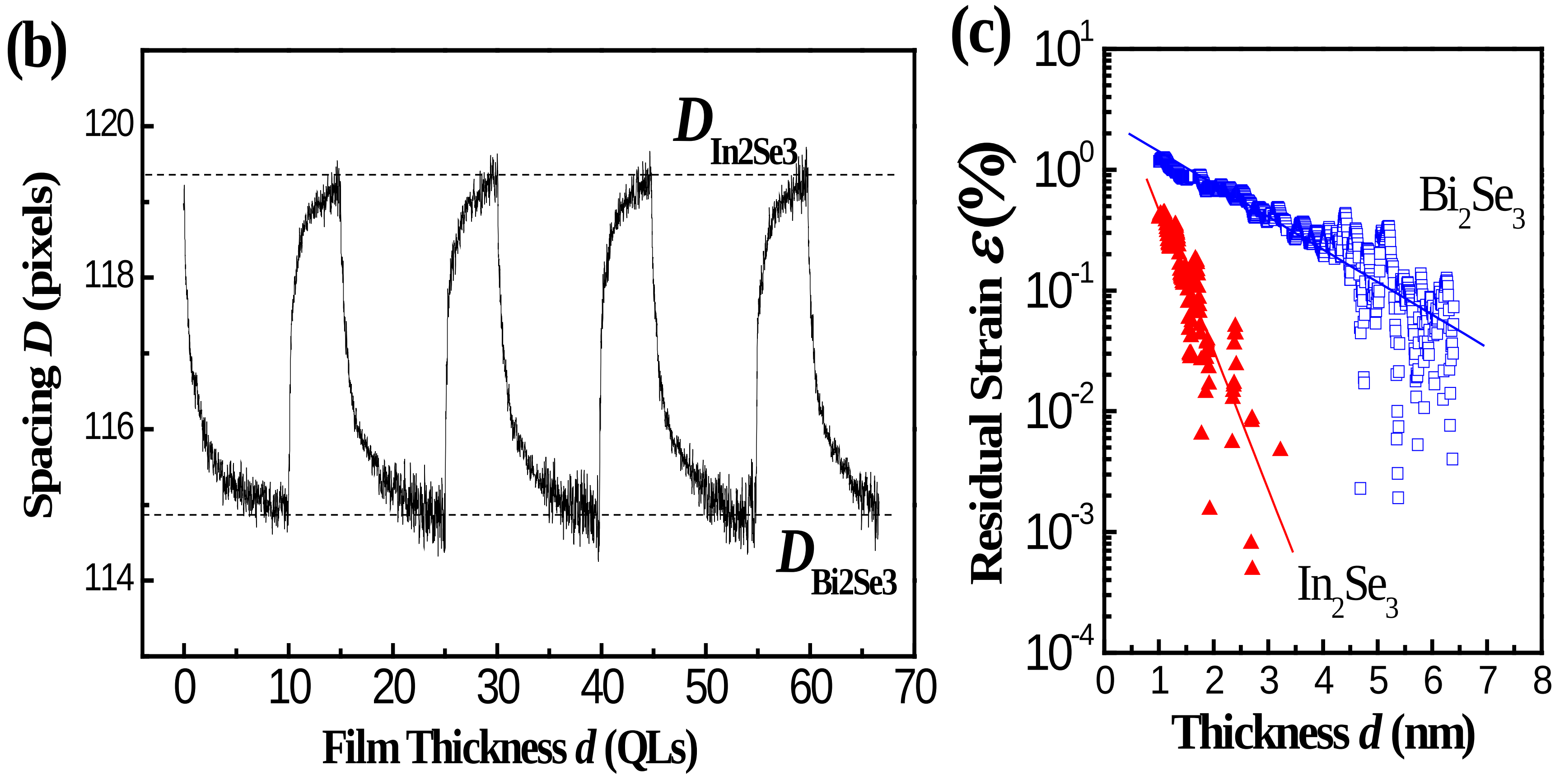

# Fig.3

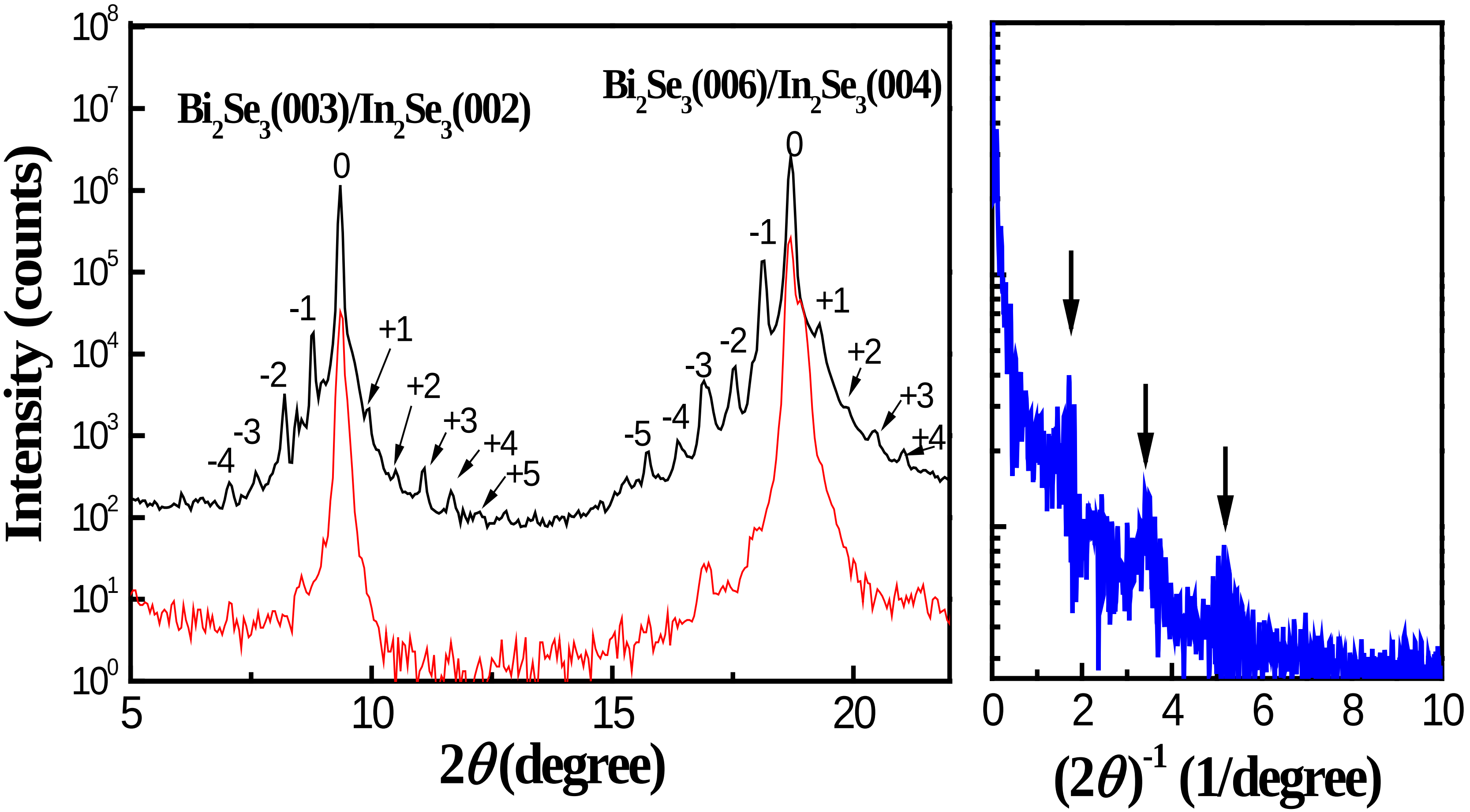

Bi2Se3(003)/In2Se3(002)
Bi2Se3(006)/In2Se3(004)
Intensity (counts)
2θ(degree)
(2θ)-1 (1/degree)

# Fig.4

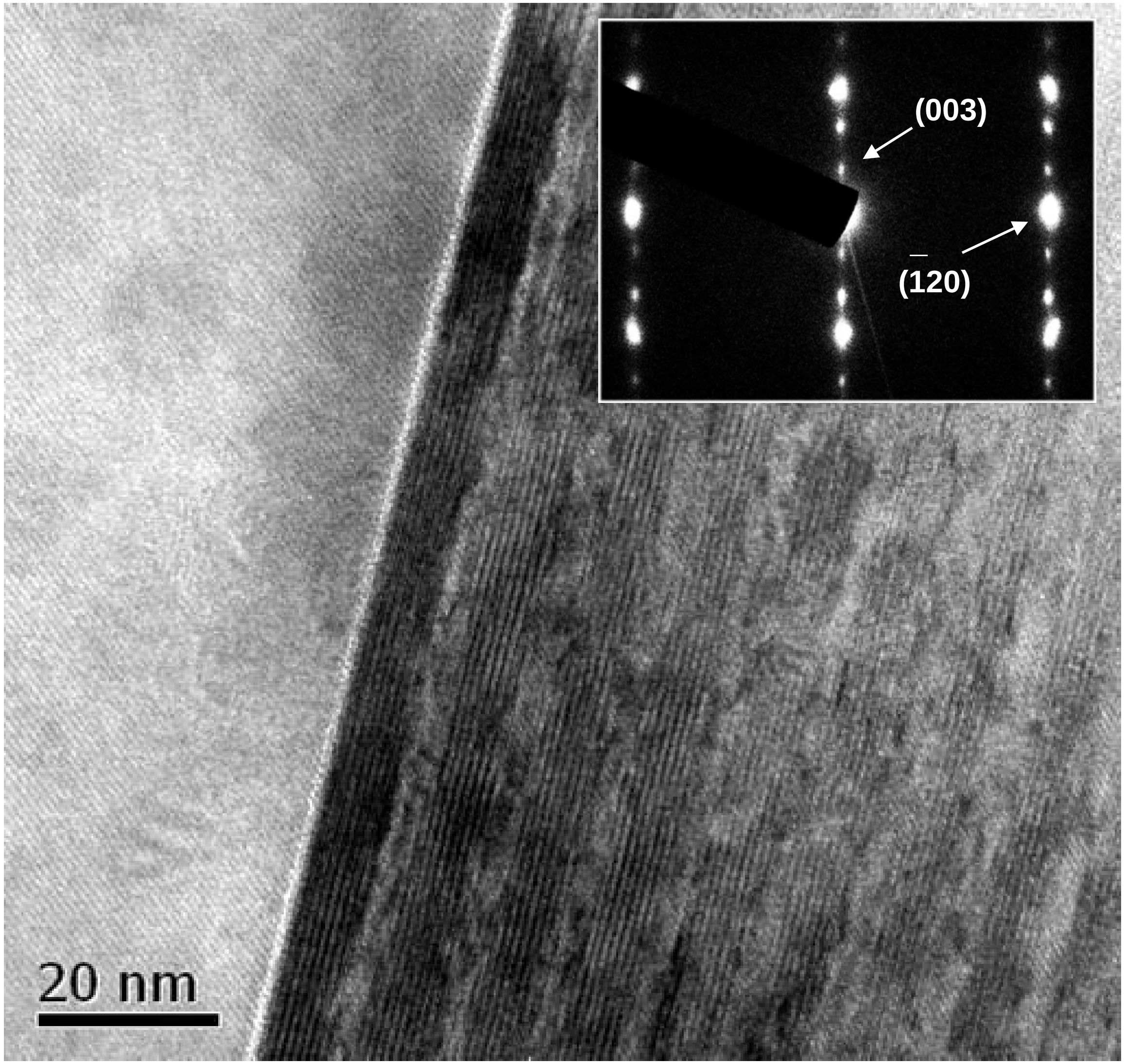